\documentclass[prx,twocolumn,twoside,preprintnumbers,superscriptaddress,nofootinbib,letterpaper]{revtex4-2}

\usepackage{amsmath,slashed}
\usepackage{graphicx, graphics, color}
\usepackage{dcolumn}
\usepackage{xspace}
\usepackage{enumerate}
\usepackage{varwidth}
\usepackage{microtype}
\usepackage{gensymb}

\usepackage{physics}
\usepackage{amssymb}
\usepackage{mathtools}
\usepackage{dsfont}
\usepackage{multirow}

\usepackage{tikz}
\usetikzlibrary{calc,arrows,arrows.meta,decorations.pathmorphing,backgrounds,positioning,fit,matrix,shapes,plotmarks,decorations.markings}
\usepackage[compat=1.1.0]{tikz-feynman}
\usepackage{contour}
\tikzfeynmanset{warn luatex=false}
\usepackage{cancel}

\usepackage[colorlinks=true,citecolor=blue,linkcolor=blue,urlcolor=blue]{hyperref}

\usepackage{orcidlink}

\newcommand{\apv}{\ensuremath{A_\mathrm{PV}}}
\newcommand{\rskin}{\ensuremath{R_\mathrm{skin}}}
\newcommand{\rchsq}{\ensuremath{R_\mathrm{ch}^2}}
\newcommand{\rch}{\ensuremath{R_\mathrm{ch}}}
\newcommand{\rchexp}{\ensuremath{R_\mathrm{ch,exp}}}
\newcommand{\rwsq}{\ensuremath{R_\mathrm{w}^2}}
\newcommand{\rw}{\ensuremath{R_\mathrm{w}}}

\newcommand{\rneut}{\ensuremath{R_n}}

\newcommand{\rprot}{\ensuremath{R_p}}

\newcommand{\MeV}{\text{MeV}}

\newcommand{\fm}{\text{fm}}
\newcommand{\ppb}{\text{ppb}}

\newcommand{\beq}{\begin{equation}}
\newcommand{\eeq}{\end{equation}}

\newcommand{\elem}[2]{\ensuremath{^{#2}\mathrm{#1}}}

\newcommand{\fmi}{\ensuremath{\text{fm}^{-1}}}

\newcommand{\nnlo}{\ensuremath{\text{N$^2$LO}}}
\newcommand{\nnnlo}{\ensuremath{\text{N$^3$LO}}}

\newcommand{\dnnlogo}{\ensuremath{\Delta\text{NNLO}_\text{GO}}}

\allowdisplaybreaks[1]

\begin{document}

\title{Ab initio calculations of parity-violating electron scattering off \texorpdfstring{$\boldsymbol{^{48}\text{Ca}}$}{} and \texorpdfstring{$\boldsymbol{^{208}\text{Pb}}$}{}}

\author{Frederic No\"el\,\orcidlink{0000-0002-7450-7213}}
\affiliation{Albert Einstein Center for Fundamental Physics, Institute for Theoretical Physics, University of Bern, Sidlerstrasse 5, 3012 Bern, Switzerland}
\author{Matthias Heinz\,\orcidlink{0000-0002-6363-0056}}
\affiliation{National Center for Computational Sciences, Oak Ridge National Laboratory, Oak Ridge, TN 37831, USA}
\affiliation{Physics Division, Oak Ridge National Laboratory, Oak Ridge, TN 37831, USA}
\author{Martin Hoferichter\,\orcidlink{0000-0003-1113-9377}}
\affiliation{Albert Einstein Center for Fundamental Physics, Institute for Theoretical Physics, University of Bern, Sidlerstrasse 5, 3012 Bern, Switzerland}
\author{Takayuki Miyagi\,\orcidlink{0000-0002-6529-4164}}
\affiliation{Center for Computational Sciences, University of Tsukuba, 1-1-1 Tennodai, Tsukuba 305-8577, Japan}
\author{Achim Schwenk\,\orcidlink{0000-0001-8027-4076}}
\affiliation{Technische Universit\"at Darmstadt, Department of Physics, 64289 Darmstadt, Germany}
\affiliation{ExtreMe Matter Institute EMMI, GSI Helmholtzzentrum f\"ur Schwerionenforschung GmbH, 64291 Darmstadt, Germany}
\affiliation{Max-Planck-Institut f\"ur Kernphysik, Saupfercheckweg 1, 69117 Heidelberg, Germany}

\begin{abstract}
 Parity-violating electron scattering off nuclei both serves as a low-energy precision probe to test electroweak interactions and allows one to access neutron distributions inside nuclei.
 It has implications for strong interactions in dense neutron-rich environments, also providing constraints for the properties of matter in neutron stars. Precision measurements are available for  $^{48}$Ca and $^{208}$Pb by the CREX and PREX collaborations, respectively, and their interpretation requires advanced nuclear-structure calculations to draw firm conclusions. We perform the first ab~initio calculations of the parity-violating asymmetry $A_\text{PV}$ based on nuclear forces from chiral effective field theory, fully including corrections due to Coulomb distortion effects. Based on these results, we critically reexamine correlation analyses employed to infer weak radii  
 and quantify the resulting tensions between ab initio and experimental results. We find that ab initio calculations prefer values of $A_\text{PV}$ slightly smaller and larger than observed for $^{48}$Ca and $^{208}$Pb, respectively, with a global significance of $1.9\sigma$.
 Using theoretically consistent inputs for charge and weak densities,
 we infer from the experimental \apv{} a neutron skin of \elem{Pb}{208} of $\rneut - \rprot = 0.187(25)(18)\,\fm$, substantially smaller than that reported by PREX~II.     
\end{abstract}

\maketitle

\section{Introduction}
Neutron skins, where the neutron density extends beyond the proton density in neutron-rich nuclei, provide a window into neutron-rich nucleonic matter.
They test our knowledge of interactions between neutrons at nuclear densities~\cite{Hebeler:2015hla}
and together with multimessenger astrophysics
constrain the equation of state of neutron star matter~\cite{Essick:2021kjb,Lattimer:2023rpe,Mammei:2023kdf,Chatziioannou:2024jsr,Mendes:2026mgc}.
However, direct experimental information on the distribution of neutrons in nuclei is scarce. One promising avenue proceeds via parity-violating electron scattering (PVES), which probes the nucleus via the weak interaction~\cite{Donnelly:1989qs}. 
The proton's weak charge is much smaller than the neutron's, so PVES gives direct insight into the weak and closely related neutron distributions of the nucleus.
Experimentally, this information is in the parity-violating asymmetry \apv{} at a momentum transfer $q=\abs{\mathbf{q}}$, constructed from the difference of elastic electron scattering cross sections
for left- and right-handed electron helicities.
In the plane-wave Born approximation, \apv{} is directly related to the weak scattering form factor $F_\text{w}(q)$ as
\begin{align}
    \apv = \frac{\big(\frac{d\sigma}{d\Omega}\big)_R - \big(\frac{d\sigma}{d\Omega}\big)_L}{\big(\frac{d\sigma}{d\Omega}\big)_R + \big(\frac{d\sigma}{d\Omega}\big)_L} \simeq -\frac{G_F q^2}{4\pi\alpha_\text{el} \sqrt{2}} \frac{Q_\text{w} F_\text{w}(q)}{Z F_\text{ch}(q)}\,, \label{eq:APV_born}
\end{align}
with the charge form factor $F_\text{ch}(q)$, the weak charge of the nucleus $Q_\text{w}$, the Fermi constant $G_F$~\cite{MuLan:2012sih}, the fine-structure constant $\alpha_\text{el}=e^2/(4\pi)$, and the nuclear charge $Z$.

Precision measurements of \apv{} are available for \elem{Ca}{48}~\cite{CREX:2022kgg} and \elem{Pb}{208}~\cite{Abrahamyan:2012gp,PREX:2021umo} from the CREX and PREX collaborations, respectively, while $Q_\text{weak}$ performed a first measurement in \elem{Al}{27}~\cite{Qweak:2021ijt}. PVES is experimentally demanding, so \apv{} is only measured at a fixed value of $q$. This implies that, contrary to parity-conserving electron scattering and charge distributions~\cite{DeVries:1987atn}, one cannot unfold the Fourier transform to extract the weak distribution encoded in $F_\text{w}$. Additionally, Eq.~\eqref{eq:APV_born} receives significant Coulomb corrections due to the distortion of the electron wave function from its interaction with the potential of the nucleus, critical for a realistic description~\cite{Horowitz:1998vv,Horowitz:1999fk,Horowitz:2012tj,Horowitz:2014bxa}. 

For these reasons,
the inference of the weak radius
or the neutron skin $\rskin = \rneut - \rprot$
(the difference between the point-neutron and point-proton radii)
from PVES relies on a correlation analysis~\cite{PREX:2021umo,CREX:2022kgg}.
Computations of nuclei with varying neutron densities show correlations between \rskin{}, $F_\mathrm{w}$, and \apv{},
which are then used to infer $F_\mathrm{w}$ and \rskin{} from the measured \apv{}.
Following this approach, a larger-than-expected neutron skin was obtained for \elem{Pb}{208}, while \rskin{} for \elem{Ca}{48} was more in line with nuclear-structure expectations~\cite{Abrahamyan:2012gp,PREX:2021umo,CREX:2022kgg}. 
The inferred neutron skins have conflicting implications for the properties of nuclear matter, triggering extensive investigations to reconcile the PREX~II and CREX measurements~\cite{Reinhard:2022inh,Yuksel:2022umn,Zhang:2022bni,Mondal:2022cva,Papakonstantinou:2022gkt,Miyatsu:2023lki,Reed:2023cap,Sammarruca:2023mxp,Zhao:2024gjz,Roca-Maza:2025vnr,Kunjipurayil:2025xss,Reed:2025ccn,Reed:2026bru,Piekarewicz:2026rfj,Reed:2026wll}.

Analyses of PVES experiments have primarily relied on calculations using energy-density functional (EDF) theory. Now ab~initio calculations of nuclei using systematically improvable many-body methods and nuclear Hamiltonians from chiral effective field theory (EFT) are able to compute the structure of both \elem{Ca}{48} and \elem{Pb}{208}~\cite{Hagen:2015yea,Hu:2021trw,Hebeler:2022aui,Arthuis:2024mnl,Bonaiti:2025bsb}.
In such calculations,
the systematic uncertainties of chiral Hamiltonians are insufficient for precision predictions,
but they may be quantified and systematically explored.
This reveals correlations between related observables, e.g., between charge radii, neutron radii, and proton, neutron, charge, and weak densities. 
Such correlations were recently analyzed to predict overlap integrals 
of lepton wave functions with proton/neutron densities
for $\mu\to e$ conversion in nuclei~\cite{Heinz:2024cwg}
and to predict the fourth-order moments of nuclear charge densities and understand their connection to neutron skins~\cite{Miyagi:2025rvx}. In general, they provide one avenue to improve nuclear structure factors for beyond-Standard-Model processes in cases in which predictions based on a single chiral Hamiltonian are not sufficiently accurate~\cite{Hoferichter:2016nvd,Hoferichter:2018acd,Payne:2019wvy,Hoferichter:2020osn,Hoferichter:2022mna}.  

\begin{table}[t]
	\renewcommand{\arraystretch}{1.3} 
	\centering 
    \begin{ruledtabular}
    \begin{tabular}{l l r r}
    & Observable & \multicolumn{1}{c}{This work} & \multicolumn{1}{c}{Experiment} \\
    \hline
    \multirow{3}{*}{\elem{Ca}{48}} & \multirow{1}{*}{\apv{} [CREX, ppb]} & $2387(12)(82)$\footnotemark[1]\footnotetext[1]{Predicted from correlation with \rchsq{}.} & \multirow{1}{*}{$2668(106)(40)$} \\
    \cline{2-4}
    & \multirow{2}{*}{$\rneut - \rprot$ [fm]} & $0.120(14)(12)$\footnotemark[2]\footnotetext[2]{Inferred from correlation with  \apv{}.} & \multirow{2}{*}{$0.121(26)(24)$} \\
    & & $0.152(1)(18)$\footnotemark[1] & \\
    \hline
    \multirow{3}{*}{\elem{Pb}{208}} & \multirow{1}{*}{\apv{} [PREX~II, ppb]} & $570.1(0.6)(5.1)$\footnotemark[1] & \multirow{1}{*}{$550(16)(8)$}\\
    \cline{2-4}
    & \multirow{2}{*}{$\rneut - \rprot$ [fm]} & $0.187(25)(18)$\footnotemark[2] & \multirow{2}{*}{$0.278(78)(12)$} \\
    & & $0.155(0)(24)$\footnotemark[1] &
	\end{tabular} 
    \end{ruledtabular}
    \caption{Predicted parity-violating asymmetries \apv{} in parts-per-billion (ppb) at the CREX and PREX~II kinematics and predicted/inferred neutron skins \rskin{} in \elem{Ca}{48} and \elem{Pb}{208} compared to results from CREX~\cite{CREX:2022kgg} and PREX~II~\cite{PREX:2021umo}.
    We present both neutron skins predicted from correlations with \rchsq{}
    and inferred from correlations with \apv{}.
    For the values given for this work,
    the first uncertainty is from the experimental value used to calibrate the correlation (either \rchsq{} or \apv{});
    the second uncertainty is 
    from the correlation itself
    due to the residual scatter of the ensemble of Hamiltonians.
    \label{tab:ResultsSummary}
    }
\end{table}

\begin{figure}[t]
    \centering
    \includegraphics[width=0.48\textwidth,trim={0.35cm 0.45cm 0.4cm 0.4cm},clip]{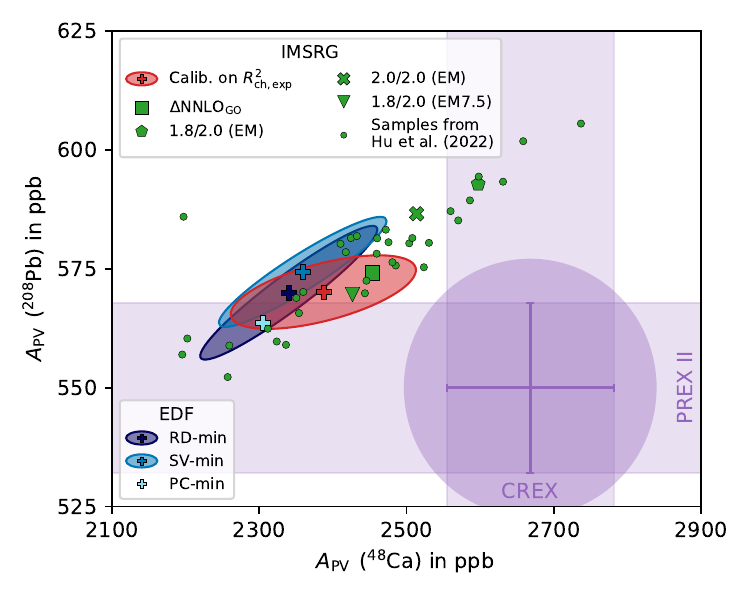}
    \caption{Ab initio \apv{} predictions from IMSRG computations of \elem{Ca}{48} and \elem{Pb}{208}
    using the chiral Hamiltonians
    1.8/2.0~(EM)~\cite{Hebeler:2010xb},
    2.0/2.0~(EM)~\cite{Hebeler:2010xb},
    \dnnlogo{}~\cite{Jiang:2020the},
    1.8/2.0~(EM7.5)~\cite{Arthuis:2024mnl},
    and 34 samples from~\textcite{Hu:2021trw} (individual points) as well as the result predicted from the correlations with the experimental $\rchsq$ (red ellipse).
    Details on the Hamiltonians can be found in the references and in App.~\ref{app:many_body}. 
    These are compared with the PREX~II and CREX measurements and predictions from EDF theory~\cite{Reinhard:2022inh}
    with the functionals SV-min~\cite{Klupfel:2008af}, RD-min~\cite{Erler:2010zh}, and PC-min~\cite{Nazarewicz:2013gda}. All ellipses correspond to 68.3\% confidence level.}
    \label{fig:NewTension}
\end{figure}

The apparent inconsistency of the PREX~II and CREX results remains a puzzle for nuclear physics.
However, still missing is an ab~initio analysis that proceeds at the level of the experimental observables \apv{} including Coulomb-distortion effects, requiring the entire charge and weak distributions for the nucleus for each chiral Hamiltonian employed. 
In this Letter, we present such an ab initio analysis.
Our approach proceeds in three steps:
\begin{enumerate}
    \item We start from an ensemble of 38 Hamiltonians~\cite{Hebeler:2010xb, Jiang:2020the, Hu:2021trw, Arthuis:2024mnl} with two- and three-nucleon interactions from chiral EFT~\cite{Epelbaum:2008ga, Machleidt:2011zz}, covering various sources of uncertainty in nuclear Hamiltonians.
    \item For each Hamiltonian, we compute the ground-state radii and charge/weak densities of \elem{Ca}{48} and \elem{Pb}{208} using the ab initio in-medium similarity renormalization group (IMSRG)~\cite{Hergert:2015awm, Heinz:2021xir}.
    We also consider experimental charge densities from electron scattering, for which we reanalyze electron scattering data in $^{208}$Pb to properly account for data and fit uncertainties of the extracted density~\cite{Noel:2024led,Noel:2024swe}.
    \item From these densities, we compute \apv{} at the kinematics of the CREX and PREX~II experiments,
    fully including Coulomb-distortion effects~\cite{phasr}.
\end{enumerate}
From this, we analyze correlations between \apv{}, nuclear radii, and neutron skins to predict \apv{} and infer \rskin{} for \elem{Ca}{48} and \elem{Pb}{208}.
We quantify the tension between measurements and nuclear-structure predictions, 
and we revisit the correlation analyses used to infer the neutron skin from \apv{}. Before giving further details on the analysis, we first summarize our main results as given in Table~\ref{tab:ResultsSummary} and Figs.~\ref{fig:NewTension} and \ref{fig:corr_APV_main}. 

\section{Main results}
\begin{figure}[t]
    \centering
    \includegraphics[width=0.96\linewidth,trim={0.37cm 0.4cm 0.42cm 0.4cm},clip]{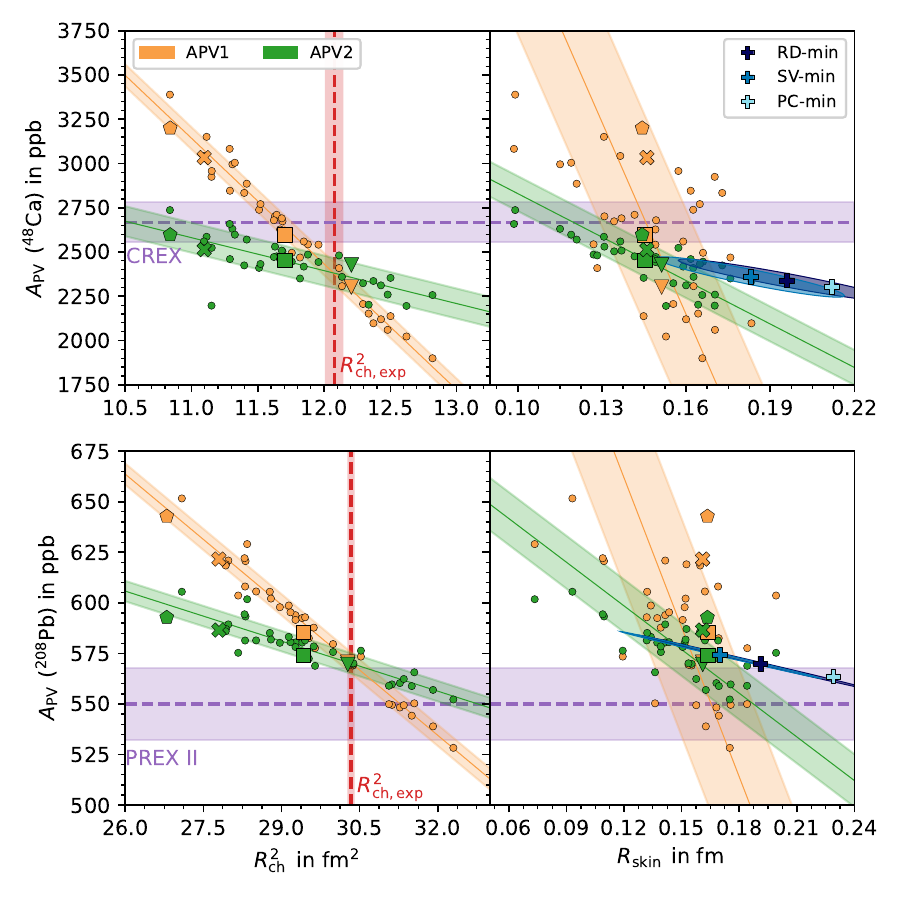}
    \caption{Ab initio predictions for \apv{} for CREX~\cite{CREX:2022kgg} (top) and PREX~II~\cite{PREX:2021umo} (bottom) versus the charge radius squared. 
For ``APV1,'' we use the charge distributions from electron scattering, while for ``APV2'' we use the theoretically predicted charge distributions for each chiral Hamiltonian (the markers are defined in Fig.~\ref{fig:NewTension}).
The vertical red band indicates the experimental charge radius squared~\cite{Noel:2024led, Sun:2025qll},
while the horizontal purple bands indicate the measured \apv{} values~\cite{PREX:2021umo, CREX:2022kgg}.
We also show results from EDFs~\cite{Reinhard:2022inh}.
}
    \label{fig:corr_APV_main}
\end{figure}

We find robust correlations between \apv{}, the charge radius squared \rchsq{}, the weak radius squared \rwsq{}, and the neutron skin \rskin{} in both \elem{Ca}{48} and \elem{Pb}{208}, see Fig.~\ref{fig:corr_APV_main}.
Using the correlation between \apv{} and \rchsq{} and a calibration on  experimental charge radii of \elem{Ca}{48} and \elem{Pb}{208} (including a reanalysis of electron scattering data~\cite{Noel:2024led}, see App.~\ref{app:charge_distribution}, and constraints from muonic atoms~\cite{Fricke:1995zz,Sun:2025qll}), we predict \apv{} for the experimental kinematics of the CREX and PREX~II experiments, summarized in Table~\ref{tab:ResultsSummary}.
For CREX, the predicted \apv{} is slightly smaller than experiment,
while for PREX~II it is slightly larger.
Both predictions show mild tension with experiment, with significances of $2.0\sigma$ and $1.1\sigma$ compared to CREX and PREX~II, respectively.
As also observed in EDF-based studies~\cite{Reinhard:2022inh},
calculations that accurately predict the charge radii of \elem{Ca}{48} and \elem{Pb}{208}
do not fully reproduce the measured \apv{} values.

This is also seen clearly in Fig.~\ref{fig:NewTension},
where we show predictions for \apv{} for both nuclei.
We show both the individual predictions for each of the 38 Hamiltonians in green
as well as the predictions from Table~\ref{tab:ResultsSummary} calibrated on the charge radii of \elem{Ca}{48} and \elem{Pb}{208} in red. 
We compare our results with a variety of EDFs~\cite{Reinhard:2022inh, Klupfel:2008af, Erler:2010zh, Nazarewicz:2013gda}.
These EDFs were selected to perform with similar quality for basic nuclear observables in different mass regions. 
Specifically, the binding energy and charge radii of \elem{Ca}{48} and \elem{Pb}{208} are predicted within about 1~MeV and 0.02~fm, respectively.
We observe a clear trend between \apv{} predictions in both nuclei, consistent with that found in EDF theory.
This trend does not overlap with the measured \apv{} values within experimental and theory uncertainties.
Our combined predictions of \apv{} for \elem{Ca}{48} and \elem{Pb}{208} show a mild global tension of $1.9\sigma$ with experiment.

We also use the correlation between \apv{} and \rskin{} in Fig.~\ref{fig:corr_APV_main}
to infer the neutron skins of \elem{Ca}{48} and \elem{Pb}{208},
given in Table~\ref{tab:ResultsSummary}.
The correlation we find is compatible with that from EDF-based analyses within uncertainties, but with a steeper slope and considerably larger uncertainty.
We infer a smaller neutron skin for \elem{Pb}{208} than was inferred by PREX~II,
but with the large uncertainties the two results are still compatible. The central value is affected 
by certain choices in the analysis, e.g., 
using a theoretical vs.\ experimental charge density or using the correlation to infer \rwsq{} vs.\ \rskin{} directly. In particular, using the experimental charge density together with calculated weak density and using correlations to infer \rw{} (in an analogous manner to the PREX~II analysis) leads to a significantly larger value for the neutron skin
than the direct inference of the neutron skin using consistent theoretical charge and weak densities.

In Table~\ref{tab:ResultsSummary}, we also show predictions for the neutron skins of \elem{Ca}{48} and \elem{Pb}{208} from a direct correlation between \rskin{} and \rchsq{} (see App.~\ref{app:further_correlations}), consistent with predictions from Refs.~\cite{Hagen:2015yea,Hu:2021trw}.
The differences in the neutron skins predicted from the correlation with \rchsq{} vs.\ those inferred from \apv{} reflect the tension between experimental \apv{} and \rchsq{} values.

While our results suggest that the tension between CREX and PREX~II for the neutron skins was overstated in the past, the measured \apv{} values are still at slight tension with constraints from the experimental charge radii of \elem{Ca}{48} and \elem{Pb}{208} at the level of significance quoted above.
Moreover, for the less stringent correlations, such as those to the neutron skin, the difference in the slope derived for the correlation of \apv{} vs.\ \rskin{} or vice versa, which differs within the statistical precision, becomes nonnegligible.
These subtleties in the correlation analysis, described in more detail below, ultimately reflect the remaining model dependence when inferring \rskin{} from the measurement of \apv{}
at a single momentum transfer, with important implications for the analysis of future experiments such as MREX~\cite{Becker:2018ggl}.

\section{Nuclear structure calculations}
We compute nuclear densities and charge radii using the ab initio IMSRG~\cite{Hergert:2015awm}.
We use an ensemble of 38 Hamiltonians with two- and three-nucleon interactions from chiral EFT to systematically explore the intrinsic uncertainty of nuclear Hamiltonians~\cite{Hebeler:2010xb, Jiang:2020the, Hu:2021trw, Arthuis:2024mnl}.
This ensemble spans multiple orders in the EFT, various regularization scales, and different approaches to fitting to data;
a notable inclusion is the set of 34 so-called nonimplausible samples from Ref.~\cite{Hu:2021trw},
which span a very conservative uncertainty of chiral EFT at next-to-next-to-leading order.
We solve the Schr\"odinger equation in the IMSRG(2) approximation, which is reliable for ground-state properties~\cite{Heinz:2021xir, Heinz:2024juw}.
For $^{48}$Ca, we use the computations from Ref.~\cite{Heinz:2024cwg}.
For our new computations of $^{208}$Pb, we use the same model-space truncations as in Refs.~\cite{Hu:2021trw, Miyagi:2025rvx},
sufficient for converged calculations of $^{208}$Pb~\cite{Hu:2021trw, Hebeler:2022aui, Arthuis:2024mnl}.
Details are provided in App.~\ref{app:many_body}.

From this, we obtain the charge radius $\rch$, the point-proton and point-neutron radii $\rprot$ and $\rneut$, the neutron skin \rskin{},
and the full intrinsic charge and weak densities
$\rho_\mathrm{ch}$ and $\rho_\mathrm{w}$ of \elem{Ca}{48} and \elem{Pb}{208} for each Hamiltonian.
$\rho_\mathrm{ch}$ and $\rho_\mathrm{w}$ enter as input into our calculations of \apv{} below,
allowing us to connect PVES with the structure of \elem{Ca}{48} and \elem{Pb}{208} and nuclear forces from chiral EFT.

\section{Charge density of \texorpdfstring{$\boldsymbol{\elem{Pb}{208}}$}{lead-208}}
To ensure a robust calibration of our correlations, 
we also reconsider the experimental charge density from elastic electron scattering. 
The commonly used collection of model-independent parameterizations of charge distributions based on electron--nucleus scattering from Ref.~\cite{DeVries:1987atn} does not provide uncertainty estimates for the truncation of the Fourier--Bessel expansion of the charge distributions.
This motivated the reanalysis of electron scattering data for isotopes relevant for $\mu\to e$ conversion in nuclei in Ref.~\cite{Noel:2024led},
which also included \elem{Ca}{48}.
This resulted in an improved charge distribution of direct relevance for PVES. 
We repeat a similar approach for \elem{Pb}{208}, see App.~\ref{app:charge_distribution} for details. 
We find that inconsistencies in the database~\cite{Bellicard:1967:2,VanNiftrik:1969lwm,Heisenberg:1969nlc,Nagao:1971mvp,Friedrich:1972iz,Friar:1973wy,Dreher:1974pqw,DeJager:1974liz,Euteneuer:1976zz,Euteneuer:1976,Frois:1977hr,Euteneuer:1977,Euteneuer:1978qw,Mazanek:1992} require a substantial inflation of uncertainties, but together with constraints from muonic atom spectroscopy~\cite{Fricke:1995zz,Sun:2025qll}
we obtain a statistically consistent and precise determination of the charge density.
Within this work, we employ this as the experimental charge distribution and corresponding experimental charge radius.

\section{Parity-violating asymmetry}
To compute \apv{},
we construct the charge and weak potentials $V_\text{ch}(r)$ and $V_\text{w}(r)$
from a given pair of charge and weak densities and the respective modifications due to $Z$-boson exchange as described in App.~\ref{app:Coulomb}.
This gives the potentials felt by left-
and right-handed electrons~\cite{Horowitz:1998vv,Horowitz:1999fk} 
\begin{equation}
    V_{L/R}(r)=V_\text{ch}(r) \mp V_\text{w}(r)\,. 
\end{equation}
From these potentials we compute the left- and right-handed differential elastic scattering cross sections $(\frac{d\sigma}{d\Omega})_{L/R}$ including Coulomb distortions using the Python package \textit{phasr}~\cite{phasr}.
In this way, we calculate \apv{} for a given initial electron energy $E$ and scattering angle $\theta$ based on inputs for the nuclear distributions. 

In practice the experiment measures and averages over a range of angles. 
Taking into account the acceptance function of the detectors $\epsilon(\theta)$, the observable becomes~\cite{PREX:2021umo,CREX:2022kgg}
\begin{equation}
\label{angular_average}
    \expval{A_\text{PV}(E)} = \frac{ \int d\theta \sin\theta ~ \epsilon(\theta) ~ A_\text{PV}(E,\theta) ~ \frac{d\sigma}{d\Omega}{(E,\theta)}  }{ \int d\theta \sin\theta ~ \epsilon(\theta) ~ \frac{d\sigma}{d\Omega}{(E,\theta)} }\,, 
\end{equation}
with the parity-conserving cross section $\frac{d\sigma}{d\Omega}$.
Angular averages for the angle $\expval{\theta}$ and momentum transfer $\expval{q^2}$ are defined in the same way.
We calculate the asymmetry using the beam energies $E$ of the respective experiments and using the provided acceptance functions, see App.~\ref{app:process_dependent_radiative_corrections}.

\section{Correlation analysis}

With this strategy, we compute \apv{} for all 38 Hamiltonians including Coulomb-distortion effects and all corrections required to match the experimental configurations in Table~\ref{tab:APV_tab}.
We explore two different approaches.
First, we employ only the theoretically predicted weak density and use the improved experimental charge density obtained in this work (see App.~\ref{app:charge_distribution}),
labeled ``APV1.''
Second, we consistently use both theoretical charge and weak densities from our computations, labeled ``APV2.''

Figure~\ref{fig:corr_APV_main} shows the resulting correlations between \apv{}, \rchsq{}, and \rskin{} for \elem{Ca}{48} (top) and \elem{Pb}{208} (bottom).
Let us first consider the correlations between \apv{} and \rchsq{} in the left panels.
While there is a significant spread in the predictions among the different Hamiltonians, we observe in both cases that \apv{} is strongly correlated with the charge radius squared.
The correlation is significantly flatter in the APV2 approach,
indicating that \apv{} is more constrained when consistent theoretical predictions for the charge and weak densities are employed.
This originates in the strong constraints on neutron skins that come from chiral EFT and the optimization of nuclear forces to nucleon--nucleon scattering data~\cite{Hagen:2015yea, Hu:2021trw, Arthuis:2024mnl}.
However, both APV1 and APV2 give consistent results for Hamiltonians that reproduce the experimental charge radius (indicated by the red band).

Following Ref.~\cite{Heinz:2024cwg},
we perform linear fits to the correlations we find.
We assign an uncertainty based on the residual scatter of the predictions for the Hamiltonians around the best fit line.
This yields the bands shown in Fig.~\ref{fig:corr_APV_main}
and the full covariance used for Fig.~\ref{fig:NewTension}.
Further details are provided in App.~\ref{app:further_correlations}.
The correlation captures the correlated systematic EFT uncertainties in nuclear Hamiltonians,
while the band quantifies the residual uncorrelated uncertainties.
The correlated uncertainties may be reduced by calibrating on experimental data.

By calibrating on the experimental charge radii of \elem{Ca}{48} and \elem{Pb}{208} (specifically using the APV2 approach),
we predict \apv{}.
The predictions for \apv{} are given in Table~\ref{tab:ResultsSummary} (and Table~\ref{tab:APV_tab} in App.~\ref{app:process_dependent_radiative_corrections}) and are consistent with the experimental values at the level of $2.0\sigma$ and $1.1\sigma$ for CREX and PREX~II, respectively. 
The differences point in opposite directions, predicting a smaller-than-observed \apv{} for \elem{Ca}{48} and a larger one for \elem{Pb}{208}. We emphasize that our analysis includes additional radiative corrections from $\gamma$--$Z$ box diagrams~\cite{Gorchtein:2008px,Gorchtein:2011mz} and vacuum polarization not yet included in the experimental analysis, see App.~\ref{app:process_dependent_radiative_corrections}, which combine to a decrease of \apv{} around $2\%$. This improves the agreement for \elem{Pb}{208} while increasing the tension in the case of \elem{Ca}{48}.

\section{Inference of neutron skins}

Next, we reassess the correlations used to extract information on radii and neutron skins in Refs.~\cite{PREX:2021umo,CREX:2022kgg}.
The correlations we find between \apv{} and \rskin{} are shown in the right panels of Fig.~\ref{fig:corr_APV_main}
and compared with results from EDFs~\cite{Reinhard:2022inh}.
In general, our correlations are compatible with the ones derived from EDFs within error, but with a larger uncertainty and a steeper slope. 

From these correlations, we explore the inference of \rskin{} from the measured \apv{} of CREX and PREX~II.
Following the correlation APV2
(using consistent theoretical charge and weak densities),
we obtain the results in Table~\ref{tab:ResultsSummary}.
For \elem{Ca}{48} our result agrees very well with that of CREX.
For \elem{Pb}{208} our result is considerably smaller  than the PREX~II central value,
but still compatible within uncertainties.
Following instead the correlation APV1 (using the experimental charge density),
our inferred value of \rskin{} is slightly smaller in \elem{Pb}{208} and slightly larger in \elem{Ca}{48}.
Clearly we observe some sensitivity to the treatment of the charge density, but we note that both approaches are consistent within uncertainties.
An exhaustive list of radii and radius differences inferred from the measured \apv{} is given in Table~\ref{tab:rsq_rskin_reverse} in the Appendix.

\begin{figure}[t]
    \centering
    \includegraphics[width=0.95\linewidth,trim= 20 20 0 10]{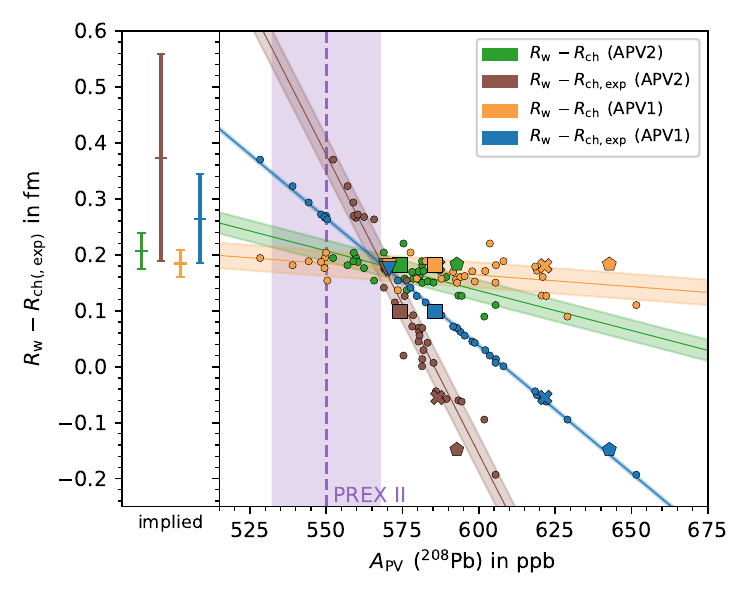}
    \caption{Correlation between \apv{} and $\rw - \rch$ radii differences for \elem{Pb}{208}
    and the values of $\rw - \rch$ implied by the \apv{} value measured by PREX~II for the different APV2 and APV1 scenarios.}
    \label{fig:rw-rch}
\end{figure}

\section{Implications for CREX and PREX~II}
Based on the sensitivity to the treatment of the charge density, we revisit the CREX and PREX~II analyses~\cite{CREX:2022kgg, PREX:2021umo}.
Both analyses use fixed charge densities from experiment in their computations of \apv{}.
The CREX analysis first varies a parameterized weak density to reproduce the measured \apv{}, yielding the form factor difference $F_\mathrm{ch}(q) - F_\mathrm{w}(q)$,
which is then used to infer \rskin{} and $\rw - \rch$ based on an EDF-based correlation.
The PREX~II analysis instead uses an EDF-based correlation between \apv{} and \rw{} to directly infer \rw{}.
Translating this into an inference of \rskin{} uses an assumed linear relationship between $\rneut - \rprot$ and $\rw - \rch$.\footnote{This relationship in Ref.~\cite{PREX:2021umo} has $\rw - \rch=0$ when $\rneut - \rprot = 0$, while we find $\rw - \rch=0.0137(3)\,\fm$ at this point.}
One way to try to emulate a similar strategy is illustrated in Fig.~\ref{fig:rw-rch}, showing, in addition to the APV2 and APV1 strategies, variants in which $\rch$ is always fixed to its experimental value. The resulting correlations imply unrealistically large values for $\rw - \rch$ (and thus of \rskin{}), albeit consistent within the very large uncertainties. This result is reminiscent of the PREX result.

Using the experimental charge density with the calculated weak density neglects the intrinsic theoretical connections between charge and weak densities.
Computations predicting larger weak radii also predict larger charge radii, and treating these consistently is important,
especially given the current state of theory:
No theoretical calculations can simultaneously reproduce the charge radii and the measured \apv{} values for \elem{Ca}{48} and \elem{Pb}{208}.
Subtracting the experimental charge radius greatly exaggerates the tension between charge radii and \apv{} values,
see Table~\ref{tab:rsq_rskin_reverse} in the Appendix for the explicit values. Ultimately, these subtleties in setting up a suitable correlation strategy reflect the challenges in inferring \rskin{} from a measurement of \apv{} at a single momentum transfer, which requires a fair amount of nuclear structure input. We believe that the analysis presented in this Letter provides a more realistic estimate of the resulting uncertainties, consistently accounting for the intrinsic connections between charge and weak densities throughout. 

\begin{table}[t]
	\renewcommand{\arraystretch}{1.3} 
	\centering
    \begin{ruledtabular}
    \begin{tabular}{l l r r} 
    \textbf{$^{\text{48}}\text{Ca}$} & Corr. & \multicolumn{1}{c}{Inferred from \apv{}} & \multicolumn{1}{c}{Experiment} \\ \colrule
    \multirow{2}{*}{\rch{}} & APV2 & $3.325(53)(52)$ & \multirow{2}{*}{$3.475(10)$~\cite{Noel:2024led}} \\
    & APV1 & $3.417(23)(13)$ &  \\
    \hline
    \multirow{2}{*}{\rw{}} & APV2 & $3.483(68)(48)$ & \multirow{2}{*}{--} \\ 
    & APV1 & $3.601(26)(10)$ &  \\ 
    \hline
    \multirow{2}{*}{$\rneut - \rprot$} & APV2 & $0.120(14)(12)$ & \multirow{2}{*}{$0.121(26)(24)$~\cite{CREX:2022kgg}} \\
    & APV1 & $0.146(3)(16)$ & \\
    \hline
    \multirow{2}{*}{$\rw - \rch$} & APV2 & $0.158(14)(13)$ & \multirow{2}{*}{$0.159(26)(23)$~\cite{CREX:2022kgg}} \\
    & APV1 & $0.184(3)(17)$ & \\
    \hline
    \multirow{2}{*}{$\rw - \rchexp$} & APV2 & $0.008(68)(48)\{10\}$ & \multirow{2}{*}{--} \\
    & APV1 & $0.126(26)(10)\{10\}$ & \\
    \hline\hline
    \textbf{$^{\text{208}}\text{Pb}$} & Corr. & \multicolumn{1}{c}{Inferred from \apv{}} & \multicolumn{1}{c}{Experiment} \\ \colrule
    \multirow{2}{*}{\rch{}} & APV2 & $5.68(16)(05)$ &  \multirow{2}{*}{$5.508(6)^*$} \\
    & APV1 & $5.588(72)(21)$ & \\
    \hline
    \multirow{2}{*}{\rw{}} & APV2 & $5.88(18)(03)$ &  \multirow{2}{*}{$5.795(82)(13)$~\cite{PREX:2021umo}} \\
    & APV1 & $5.772(79)(5)$ &  \\ 
    \hline
    \multirow{2}{*}{$\rneut - \rprot$} & APV2 & $0.187(25)(18)$ & \multirow{2}{*}{$0.278(78)(12)$~\cite{PREX:2021umo}} \\
    & APV1 & $0.165(7)(22)$ & \\
    \hline
    \multirow{2}{*}{$\rw - \rch$} & APV2 & $0.207(25)(19)$ & \multirow{2}{*}{--} \\
    & APV1 & $0.184(7)(23)$ & \\
    \hline
    \multirow{2}{*}{$\rw - \rchexp$} & APV2 & $0.37(18)(3)\{1\}$ & \multirow{2}{*}{--} \\ 
	& APV1 & $0.264(79)(5)\{13\}$ & 
	\end{tabular}
    \end{ruledtabular}
    \caption{Radii and skin thickness in fm 
   inferred from correlations with \apv{} 
   compared to values from experiment.
   We indicate the new charge radius from the reanalysis of electron scattering data for \elem{Pb}{208} in this work by an asterisk.
   For the inferred radii, the first uncertainty is due to the measured value of \apv{} and the second one stems from the uncertainty in the correlation. 
   For $\rw - \rchexp$, we indicate the additional uncertainty from $\rchexp$ in curly brackets.} 
	\label{tab:rsq_rskin_reverse}
\end{table}

\begin{acknowledgments}
We thank C.~J.\ Horowitz, B.~T.\ Reed, P.-G.\ Reinhard, and X.~Roca-Maza for valuable discussions and correspondence, as well as B.-L.\ Hoid and N.~S.~Oreshkina for help in retrieving Refs.~\cite{Euteneuer:1976,Mazanek:1992}.  
Financial support by the Swiss National Science Foundation (Project No.\ TMCG-2\_213690) is gratefully acknowledged.
This work was supported 
by the U.S.~Department of Energy, Office of Science, Office of Advanced Scientific Computing Research and Office of Nuclear Physics, Scientific Discovery through Advanced Computing (SciDAC) program (SciDAC-5 NUCLEI); by the Laboratory Directed Research and Development Program of Oak Ridge National Laboratory, managed by UT-Battelle, LLC, for the U.S.\ Department of Energy;
by the JST ERATO Grant No.~JPMJER2304, Japan;
by JSPS KAKENHI Grant Numbers 25K07294, 25K00995, 25K07330, and 26H01394;
and by the European Research Council (ERC) under the European Union's Horizon 2020 research and innovation programme (Grant Agreement No.~101020842).
An award of computer time was provided by the INCITE program.
This research used resources of the Oak Ridge Leadership Computing Facility at the Oak Ridge National Laboratory, which is supported by the Advanced Scientific Computing Research programs in the Office of Science of the U.S. Department of Energy under Contract No. DE-AC05-00OR22725. 
We also gratefully acknowledge the Gauss Centre for Supercomputing e.V. (www.gauss-centre.eu) for providing computing time on the GCS Supercomputer JUWELS at Jülich Supercomputing Centre (JSC). 
\end{acknowledgments}

\appendix

\section{Charge distribution of \texorpdfstring{$\boldsymbol{\elem{Pb}{208}}$}{lead-208}}
\label{app:charge_distribution}

Elastic electron scattering off \elem{Pb}{208} was studied in detail in Refs.~\cite{Bellicard:1967:2,VanNiftrik:1969lwm,Heisenberg:1969nlc,Nagao:1971mvp,Friedrich:1972iz,Friar:1973wy,Dreher:1974pqw,DeJager:1974liz,Euteneuer:1976zz,Euteneuer:1976,Frois:1977hr,Euteneuer:1977,Euteneuer:1978qw,Mazanek:1992}, but tensions in the database were never resolved, leading to a situation in which two parameterizations of the charge distribution were quoted in Ref.~\cite{DeVries:1987atn}. We have performed a comprehensive review of the literature and the available datasets, the most relevant of which emerge from the Stanford~\cite{Heisenberg:1969nlc} (and later updates as documented in Ref.~\cite{Friar:1973wy}) and Mainz~\cite{Euteneuer:1976zz,Euteneuer:1976,Euteneuer:1978qw} measurements, the latter being revised in Ref.~\cite {Mazanek:1992}. However, comparing the two charge distributions from Ref.~\cite{DeVries:1987atn} with the underlying datasets (taken as the most recent tables from Refs.~\cite{Friar:1973wy,Mazanek:1992}, respectively), we observe that the internal inconsistencies in the datasets are at least as large as the discrepancies between them.
Accordingly, we conclude that a significant $\chi^2$ inflation is required in either case to account for likely underestimated systematic uncertainties, and this situation changes very little when performing a combined fit. Therefore, we follow the global fitting strategy as described in Ref.~\cite{Noel:2024led}, including quantifying truncation uncertainties of the Fourier--Bessel expansion. 
 
\begin{figure}[t]%
    \centering
    \includegraphics[width=0.95\linewidth]{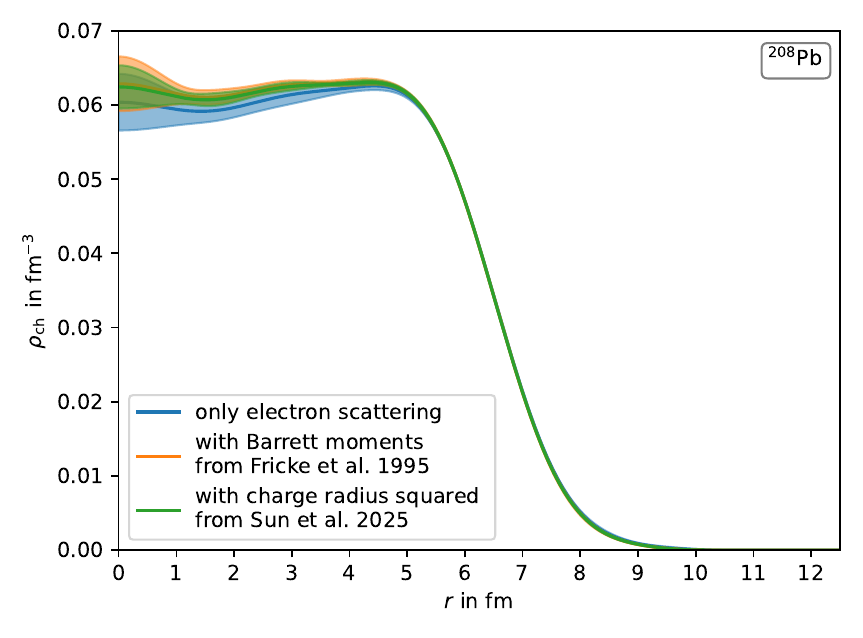} 
    \caption{Charge distributions for \elem{Pb}{208}, either from electron scattering alone (blue), constrained by the Barrett moment~\cite{Fricke:1995zz} (orange), or the charge radius~\cite{Sun:2025qll} (green).}
    \label{fig:rho_Pb208}
\end{figure}

We distinguish three scenarios: A pure electron scattering fit and two fits with different inputs from muonic-atom spectroscopy. We consider first a fit including Barrett moments from Ref.~\cite{Fricke:1995zz} as an additional constraint. Recently, the muonic-atom spectroscopy measurements of \elem{Pb}{208} were reanalyzed in Ref.~\cite{Sun:2025qll}, resulting in a refined value for the charge radius. 
We consider a fit including this radius value (instead of the Barrett moment) as an additional constraint as a third scenario. The resulting charge distributions are shown in Fig.~\ref{fig:rho_Pb208}, available as ancillary files in the same conventions as provided in the context of Ref.~\cite{Noel:2024led}. 
The predicted radii of the three parameterizations are 
\begin{align}
\rch&=5.532(15)\mqty{(20)[25]\\(\substack{+29\\-6})[32]}\,\fm && \text{(scattering only)}\,,\\
\rch&=5.503(2)\mqty{(10)[10]\\(\substack{+2\\-2})[3]}\,\fm && \text{(Barrett moment~\cite{Fricke:1995zz})}\,,\notag\\
\rch&=5.5077(36)\mqty{(52)[63]\\(\substack{+8\\-7})[37]}\,\fm && \text{(charge radius~\cite{Sun:2025qll}})\,,\notag
\end{align}
respectively, using the same strategies for statistical and systematic uncertainties as in Ref.~\cite{Noel:2024led} (the errors in square brackets give the quadratic sum). The first common error refers to statistical uncertainties, while the upper and lower second errors indicate two ways of estimating the systematic uncertainties (we use the upper one in this work as the more realistic choice in this case).

Figure~\ref{fig:rho_Pb208} illustrates that the uncertainties in the charge distribution are significant, especially near the center of the nucleus and if  solely electron scattering data are considered. Including the constraints from muonic-atom spectroscopy, the errors tend to decrease and the Barrett moment or charge radius is essentially fixed by the additional input constraint. 
Hence, a precise extraction of a charge distribution based on electron-scattering datasets relies on the input from muon spectroscopy to constrain the charge radius. On the contrary,  extractions of the charge radius from muonic atoms typically involve some model assumptions (such as a two- or three-parameter Fermi distribution~\cite{Sun:2025qll}), which emphasizes the importance of cross-validation between the two methods. To do so beyond the current uncertainties would require improved measurements of parity-conserving electron scattering to resolve the inconsistencies in the database that dominate the final uncertainty budget. 

\begin{figure}[t]%
    \centering
    \includegraphics[width=0.95\linewidth]{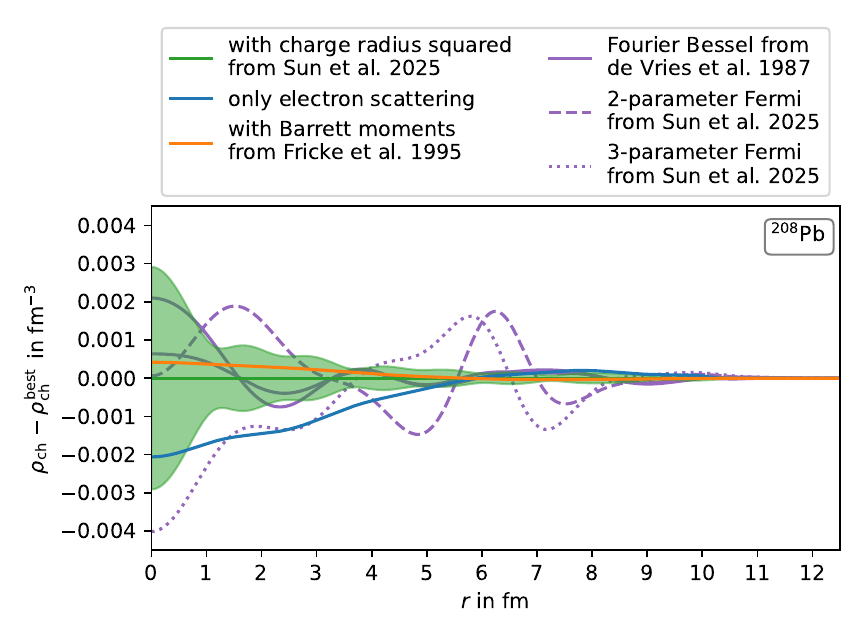} 
    \caption{Comparison of the different \elem{Pb}{208} charge distributions from Fig.~\ref{fig:rho_Pb208}, the Fourier--Bessel 1 parameterization from Ref.~\cite{DeVries:1987atn}, and the Fermi distributions from Ref.~\cite{Sun:2025qll}.}
    \label{fig:rho_Pb208_diff}
\end{figure}%
 
After the $\chi^2$ inflation, we do not observe any remaining tension between the different fit strategies and choose the most precise one with the constraint from Ref.~\cite{Sun:2025qll} as our final result, employed throughout the rest of this work. If we compare this result to the parameterizations from Ref.~\cite{DeVries:1987atn}, we find that our results are within uncertainties of both variants presented therein, both of which are also based on Fourier--Bessel expansions. 
Meanwhile, comparing the two- and three-parameter Fermi distributions 
with the same charge radius used in Ref.~\cite{Sun:2025qll}, we find significant differences in particular in the slope of the charge distribution in the surface at $r=(5$--$8)\fm$, which the simple parameterizations are unable to reproduce accurately. Both comparisons are shown in Fig.~\ref{fig:rho_Pb208_diff}. 

\section{Process-dependent radiative corrections}
\label{app:process_dependent_radiative_corrections}

\begin{table}[t]
	\renewcommand{\arraystretch}{1.3}
	\centering 
    \begin{tabular}{l l r r r r} 
	\toprule 
    & & This work & Reference \\ \colrule
     & $E_\text{exp}$ \ $[\text{GeV}]$ & -- & $2.182(5)$\\
    \elem{Ca}{48}& $\expval{\theta}$ \ $[\degree]$ & $4.507(3)$ & $4.51(2)$\\
    \cite{CREX:2022kgg}  & $q_\text{exp}^2=\expval{q^2}$ \ $[\text{GeV}^2]$ & $0.02961(4)$ & $0.0297(2)$ \\
    & $\expval{A_\text{PV}(E_\text{exp})}$ \ $[\ppb]$ & $2387(12)(82)$ & $2668(106)(40)$\\\colrule
     & $E_\text{exp}$ \ $[\text{GeV}]$ & -- & $1.060$\\
    \elem{Pb}{208}& $\expval{\theta}$ \ $[\degree]$ & $5.0395(7)$ & $\sim 5$\\
   \cite{Abrahamyan:2012gp} & $q_\text{exp}^2=\expval{q^2}$ \ $[\text{GeV}^2]$ & $0.008785(3)$ & $0.00880(11)$ \\
    & $\expval{A_\text{PV}(E_\text{exp})}$ \ $[\ppb]$ & $703.2(1.1)(8.0)$ & $656(60)(14)$\\\colrule
     & $E_\text{exp}$ \ $[\text{GeV}]$ & --  & $0.953$ \\
    \multirow{2}{*}{\elem{Pb}{208}} & \multirow{2}{*}{$\expval{\theta}$ \ $[\degree]$}  & \multirow{2}{*}{$4.6865(6)$} & $4.67(2)$\\
    & & & $4.71(2)$ & \\
    \cite{PREX:2021umo} & $q_\text{exp}^2=\expval{q^2}$ \ $[\text{GeV}^2]$ & $0.006147(2)$ & $0.00616$\\
    & $\expval{A_\text{PV}(E_\text{exp})}$ \ $[\ppb]$ & $570.1(0.6)(5.1)$ & $550(16)(8)$\\\colrule
     & $E_\text{exp}$ \ $[\text{GeV}]$ & -- & $1.157$ \\
    \multirow{2}{*}{\elem{Al}{27}} & $\expval{\theta}$ \ $[\degree]$ & -- & $7.61(2)$\\
    & $q_\text{exp}^2=\expval{q^2}$ \ $[\text{GeV}^2]$ & -- & $0.02357(10)$\\
    \cite{Qweak:2021ijt} & $\expval{A_\text{PV}(E_\text{exp})}$ \ $[\ppb]$ & -- & $2160(110)(160)$ \\
    & $A_\text{PV}(E_\text{exp},\expval{\theta})$ $[\ppb]$ & $2070(0)(12)$ & -- \\
    \botrule 
	\end{tabular} 
    \caption{Experimental specifications for the different PVES experiments~\cite{PREX:2021umo,Abrahamyan:2012gp,CREX:2022kgg,Qweak:2021ijt} (with some parameters as specified in Refs.~\cite{Reinhard:2022inh,Roca-Maza:2025vnr}), and comparison to our results (the first error is propagated from the reference value for the charge radius, the second one derives from the correlation). The angular averages for the angle and momentum transfer only depend on the charge distribution and not the weak physics. $Q_\text{weak}$~\cite{Qweak:2021ijt} do not quote an acceptance function, and we therefore use the quoted angular averages as given.} 
	\label{tab:APV_tab}
\end{table}

The calculation of radiative corrections needs to follow the experimental specifications, summarized in Table~\ref{tab:APV_tab}.
Coulomb corrections constitute by far the most important class of radiative corrections, but at increased levels of precision additional effects arise, as most evident in the case of the weak charge of the proton~\cite{Qweak:2018tjf}. First, short-distance radiative corrections modify the tree-level values of the weak charges $Q_\text{w}^p$ and $Q_\text{w}^n$~\cite{Erler:2013xha}, so that these quantities become process dependent~\cite{Crivellin:2021bkd,Abdullah:2022zue}. In addition, nonperturbative corrections from $\gamma$--$Z$ box diagrams even depend on the specific kinematic setting of the reaction~\cite{Gorchtein:2008px,Gorchtein:2011mz}, so that, to ensure consistency with the measurements, we employ the values of the weak charges as quoted in Refs.~\cite{PREX:2021umo,CREX:2022kgg},
\begin{equation}
\label{Qw_final}
    Q_\text{w}^\text{CREX} = -26.0(1)\,,\qquad  
    Q_\text{w}^\text{PREX} = -117.9(3)\,,
\end{equation}
since these radiative corrections have not been removed from the quoted values of \apv. Further radiative corrections end up being small~\cite{Roca-Maza:2025vnr,Reed:2026bru,Reed:2026wll}, leaving a sub-percent reduction of \apv{} due to vacuum-polarization effects:
\begin{equation}
    \expval{A_\text{PV}}_\text{exp} = \frac{\expval{A_\text{PV}}}{1 + \epsilon_\text{VP}(q^2)}\,,
\end{equation}
where $\expval{A_\text{PV}}_\text{exp}$ is the measured parity-violating asymmetry, $ \expval{A_\text{PV}}$ is the predicted one using the weak charges~\eqref{Qw_final} (including Coulomb-distortion effects),  and the vacuum-polarization correction reads~\cite{Milstein:2004xh,Schwartz:2014sze,Reed:2026bru} 
\begin{equation}
 \epsilon_\text{VP}(q^2)=\frac{\alpha_\text{el}(m_e^2)}{3\pi} \bigg(\log\frac{q^2}{m_e^2}-\frac{5}{3}\bigg)\,,   
\end{equation}
with momentum transfers as given in Table~\ref{tab:APV_tab}.

\section{Inclusion of Coulomb corrections}
\label{app:Coulomb}

 The charge potential $V_\text{ch}$ and weak potential $V_\text{w}$
 that enter the numerical solution of the Dirac equation
 are given as  
\begin{align}
    V_\text{ch}(r) &= - \sqrt{4\pi\alpha_\text{el}} \int_r^\infty dx\, E(x)\,, \label{eq:Vch}\\
    E(r) &= \frac{\sqrt{4\pi \alpha_\text{el}}}{r^2} \int_0^r dx\, x^2 \rho_\text{ch}(x)\,, \label{eq:El}\\
    V_\text{w}(r) &= \frac{G_F}{2\sqrt{2}} \rho_\text{w} (r)\,, \label{eq:Vw}
\end{align}
based on the charge distribution $\rho_\text{ch}$ and the weak distribution $\rho_\text{w}$, respectively.
These are defined as the Fourier transforms of the charge form factor $F_\text{ch}$ and the weak form factor $F_\text{w}$ according to
\begin{align}
    \rho_\text{ch}(r)&=\frac{1}{2\pi^2}\int dq\, q^2\,j_0(qr) \, Z F_\text{ch}(q)\,, \label{eq:rho_ch} \\
    \rho_\text{w}(r)&=\frac{1}{2\pi^2}\int dq\, q^2\,j_0(qr) \, Q_\text{w} F_\text{w}(q)\,, \label{eq:rho_w}
\end{align}
where $j_0(x)=\frac{\sin x}{x}$ denotes the lowest spherical Bessel function.
The charge and weak form factors can be calculated as combinations of nuclear structure functions of different nuclear multipoles according to~\cite{Hoferichter:2020osn} 
\begin{widetext}
\begin{align}
    Z F_\text{ch}(q)
    &=  \bigg(1 - \frac{\expval{r_p^2}}{6} q^2 - \frac{1}{8 m_N^2} q^2\bigg) \mathcal{F}^{M}_p(q) - \frac{\expval{r_n^2}}{6} q^2 \mathcal{F}^{M}_n(q) 
  +\frac{1 + 2 \kappa_p}{4 m_N^2} q^2 \mathcal{F}^{\Phi^{\prime\prime}}_p(q) + \frac{ 2 \kappa_n}{4 m_N^2} q^2 \mathcal{F}^{\Phi^{\prime\prime}}_n(q) + \mathcal{O}\big(q^4\big)\,, \label{eq:F_ch}\\
    Q_\text{w} F_\text{w}(q) 
    &=\bigg[ Q_\text{w}^p \bigg(1 -\frac{\expval{r_p^2}}{6} q^2 - \frac{1}{8 m_N^2} q^2\bigg) - Q_\text{w}^n \frac{\expval{r_n^2} + \expval{r_{s,N}^2}}{6} q^2 \bigg] \mathcal{F}^{M}_{p}(q) 
+ \frac{Q_\text{w}^p (1 + 2 \kappa_p) + 2 Q_\text{w}^n (\kappa_n + \kappa_{s,N}) }{4 m_N^2} q^2 \mathcal{F}^{\Phi^{\prime\prime}}_{p}(q)\notag\\
&+\bigg[Q_\text{w}^n \bigg( 1 - \frac{\expval{r_n^2} + \expval{r_{s,N}^2}}{6} q^2 - \frac{1}{8 m_N^2} q^2\bigg) - Q_\text{w}^p \frac{\expval{r_n^2}}{6} q^2 \bigg] \mathcal{F}^{M}_{n}(q) 
+\frac{Q_\text{w}^n (1 + 2 \kappa_p + 2 \kappa_{s,N}) + 2 Q_\text{w}^p \kappa_n }{4 m_N^2} q^2 \mathcal{F}^{\Phi^{\prime\prime}}_n(q)\,,  \label{eq:F_w}
\end{align}
\end{widetext}
where the weak charges at tree level are $Q_\text{w}^p=1-4\sin^2\theta_\text{w}$, $Q_\text{w}^n=-1$, $Q_\text{w} = Z Q^\text{w}_p + N Q^\text{w}_n$.
Throughout, we follow the conventions of Ref.~\cite{ParticleDataGroup:2026} for the electroweak parameters, including the weak mixing angle $\theta_\text{w}$. The constants appearing in Eqs.~\eqref{eq:F_ch} and~\eqref{eq:F_w} are set to the values in Table~\ref{tab:nuc_constants}. Finally, with the  structure functions of the multipoles $M$ and $\Phi''$ calculated with the IMSRG, the charge and weak distributions follow by Fourier transform according to Eqs.~\eqref{eq:rho_ch} and \eqref{eq:rho_w}, which ultimately gives the charge and weak potentials according to Eqs.~\eqref{eq:Vch} and \eqref{eq:Vw}.

\begin{table}[t]
    \centering
    \renewcommand{\arraystretch}{1.3} 
	\centering 
    \begin{tabular}{l l r r} 
	\toprule 
        $\expval{r_p^2}$ & $[\text{fm}^2]$ & $0.7071(7)$ & \cite{Antognini:2013txn}\\
        $\expval{r_n^2}$ & $[\text{fm}^2]$ & $-0.1155(17)$ & \cite{ParticleDataGroup:2026,Koester:1995nx,Kopecky:1997rw}\\
        $ \expval{r_{s,N}^2}$ & $[\text{fm}^2]$ & $-0.0048(6)$ & \cite{Djukanovic:2019jtp,Alexandrou:2019olr}\\
        $\kappa_p$ & & $1.79284734462(82)$ & \cite{ParticleDataGroup:2026,Tiesinga:2021myr,Schneider:2017lff}\\
        $\kappa_n$ & & $-1.91304273(45)$ & \cite{ParticleDataGroup:2026,Tiesinga:2021myr}\\
        $\kappa_{s,N}$ & & $-0.017(4)$ & \cite{Djukanovic:2019jtp,Alexandrou:2019olr}\\
    \botrule
    \end{tabular}
    \caption{Input values for the constants required for the evaluation of $F_\text{ch}$ and $F_\text{w}$ from the nuclear $M$ and $\Phi''$ responses.}
    \label{tab:nuc_constants}
\end{table}

The potentials are most easily derived from the Feynman rules in momentum space, which gives 
\begin{align}
    V_\text{ch}(q)= -\frac{4\pi\alpha_\text{el}}{q^2} Z F_\text{ch}(q)\,,\qquad
    V_\text{w}(q)= \frac{G_F}{2\sqrt{2}} Q_\text{w} F_\text{w}(q)\,, 
\end{align}
where we already dropped the parity-conserving contribution from $Z$-boson exchange as well as subleading spin-dependent contributions for nuclei with $J\neq0$. The position-space form then follows by Fourier transform 
\begin{align}
   V_\text{ch}(r) &= -\frac{(4\pi)^2\alpha_\text{el}}{(2\pi)^3} \int dq\, q^2 j_0(q r) \frac{Z F_\text{ch}(q)}{q^2} \nonumber\\
   & = -8\alpha_\text{el} \int_0^\infty dx\, x^2 \rho_\text{ch}(x) \frac{\pi}{2 r_>} \notag\\
   & = - \sqrt{4\pi\alpha_\text{el}} \int_r^\infty dx\, E(x)\,, \\
   V_\text{w}(r)&= \frac{G_F}{2\sqrt{2}}  \frac{4\pi}{(2\pi)^3} \int dq\, q^2 j_0(q r) Q_\text{w} F_\text{w}(q) \nonumber\\
   &= \frac{G_F}{2\sqrt{2}} \rho_\text{w}(r)\,, 
\end{align}
where  for the charge potential we used the identity
\begin{equation}
    \int_0^\infty dq\, j_0(q r) j_0(q x) = \frac{\pi}{2} \frac{1}{r_>}\,, \qquad r_>=\text{max}(r,x)\,,  
\end{equation}
as well as the definition~\eqref{eq:El} for the electric field, leading to the potentials in Eqs.~\eqref{eq:Vch} and \eqref{eq:Vw}.

\section{IMSRG calculations}
\label{app:many_body}

Our IMSRG calculations start from nuclear Hamiltonians with two-nucleon (NN) and three-nucleon (3N) interactions from chiral EFT.
We use an ensemble of 38 Hamiltonians to systematically explore the intrinsic uncertainty of nuclear Hamiltonians~\cite{Hebeler:2010xb, Jiang:2020the, Hu:2021trw, Arthuis:2024mnl}.
These include interactions at next-to-next-to-leading order (\nnlo{}) and next-to-next-to-next-to-leading order (\nnnlo{}) in chiral EFT
and interactions including explicit $\Delta$-isobar degrees of freedom in the EFT.

The Hamiltonians 1.8/2.0~(EM) and 2.0/2.0~(EM)~\cite{Hebeler:2010xb} follow the same construction.
They start from the \nnnlo{} NN potential with a cutoff of $500\,\MeV$ from Ref.~\cite{Entem:2003ft},
optimized to NN scattering data.
The NN potential is then evolved to lower resolution scales $\lambda$ using the similarity renormalization group~\cite{Bogner:2006pc},
with $\lambda=1.8\,\fmi$ for 1.8/2.0~(EM) and $\lambda=2.0\,\fmi$ for 2.0/2.0~(EM).
Subsequently the 3N potential is constructed at \nnlo{} with a low cutoff of $2.0\,\fmi$.
For both Hamiltonians, the long-range 3N couplings $c_i$ are taken from Ref.~\cite{Entem:2003ft}
and short-range 3N couplings $c_D$, $c_E$
are fit to the $^3$H binding energy and $^4$He point-proton radius.
The Hamiltonian 1.8/2.0~(EM7.5)~\cite{Arthuis:2024mnl} is constructed the same way as 1.8/2.0~(EM) except short-range 3N couplings $c_D$, $c_E$
are optimized to the $^3$H binding energy and the $^{16}$O ground-state energy and charge radius.

The \dnnlogo{} Hamiltonian~\cite{Jiang:2020the} is instead constructed
with NN and 3N potentials at \nnlo{} in chiral EFT with explicit inclusion of $\Delta$ isobars.
The pion--nucleon couplings that also enter the long-range parts of the 3N potential were taken from Ref.~\cite{Siemens:2016jwj} (determined based on Refs.~\cite{Hoferichter:2015tha,Hoferichter:2015hva}).
The remaining NN and 3N couplings were optimized to low-energy NN scattering phase shifts,
properties of nuclei with $A=2$, $3$, and $4$
and properties of infinite nuclear matter.

The 34 samples from \textcite{Hu:2021trw} are generated somewhat differently.
They also consist of NN and 3N potentials at \nnlo{} in chiral EFT with explicit $\Delta$ isobars.
However, instead of optimizing the couplings to data,
a history matching approach tests millions of parameterizations,
each with varying values for all couplings.
The pion--nucleon couplings are sampled from a prior based on Ref.~\cite{Siemens:2016jwj}
while the short-range NN and 3N couplings are sampled over broad intervals.
The parameterizations are then tested against experimental data,
including NN scattering data and
ground-state properties of nuclei with $A=2$, $3$, $4$, and $16$,
and excluded if they fail an implausibility check.
Critically, this implausibility check also accounts for the EFT uncertainty at \nnlo{}.
34 parameterizations survived this procedure, serving as the Hamiltonians we use in this work.
Because of the conservative construction,
this ensemble of 34 Hamiltonians spans a large uncertainty of chiral EFT at \nnlo{},
making it very valuable for our analysis.

\begin{figure}[t]
    \centering
    \includegraphics[width=0.95\linewidth,trim= 0 0 0 0,clip]{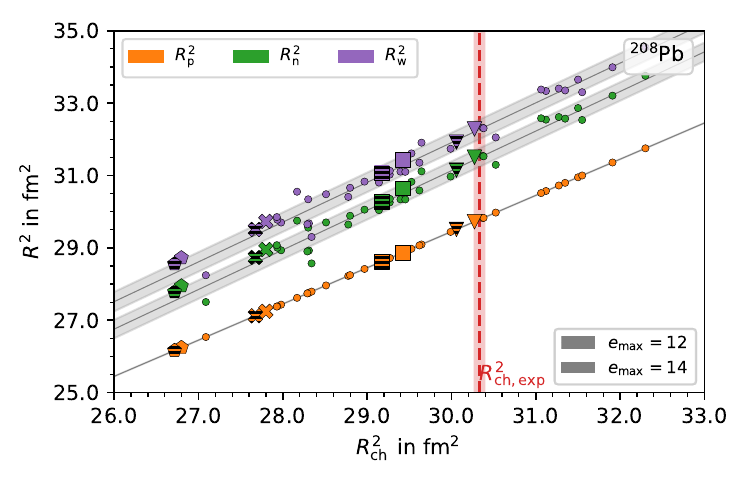}
    \caption{Radii correlations for different many-body truncations in the case of \elem{Pb}{208}.}
    \label{fig:corr_truncation}
\end{figure}

Our IMSRG calculations solve the Schr\"odinger equation for the intrinsic Hamiltonian
\begin{equation}
    H = T_\mathrm{int} +V_\mathrm{NN} + V_\mathrm{3N}\,,
\end{equation}
with the intrinsic kinetic energy $T_\mathrm{int}$
(with the center-of-mass removed),
the NN potential $V_\mathrm{NN}$,
and the 3N potential $V_\mathrm{3N}$.
We expand our calculations in a spherical harmonic oscillator (HO) basis of 15 major shells.
For $^{48}$Ca, we use the radii and densities computed in Ref.~\cite{Heinz:2024cwg}, which used an HO basis frequency $\hbar\omega = 16\,\MeV$.
For our new computations of the radii and densities of $^{208}$Pb, we use an HO basis frequency $\hbar\omega = 12\,\MeV$.
Three-body matrix elements are truncated with $E_\mathrm{3max} = 28$~\cite{Miyagi:2023qce}.
Our calculations start from a Hartree--Fock reference state
and solve the IMSRG equations at the normal-ordered two-body level, the IMSRG(2), which is reliable for ground-state properties~\cite{Heinz:2021xir, Heinz:2024juw}.
For \elem{Ca}{48},
we use the Magnus expansion approach~\cite{Morris:2015yna} without splitting of the Magnus operator.
For \elem{Pb}{208}, we instead split the Magnus operator following Refs.~\cite{Stroberg:2024zyg, Heinz:2026srm}.
The differences between these approaches were shown to be very small for IMSRG calculations of \elem{Ca}{48}~\cite{Heinz:2026srm},
and the split Magnus approach is generally expected to be more reliable, which is why we use it for our computations for \elem{Pb}{208}.

\begin{figure}[t!]
    \centering
    \includegraphics[width=0.95\linewidth,trim= 0 10 0 0,clip]{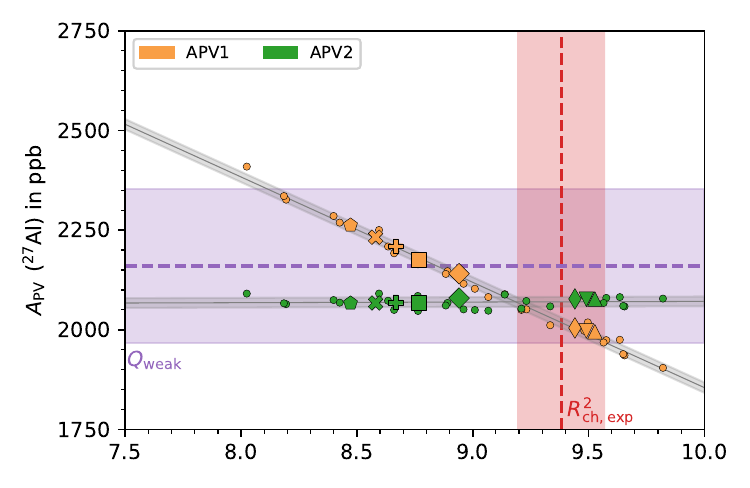}
    \includegraphics[width=0.95\linewidth,trim= 0 0 0 10,clip]{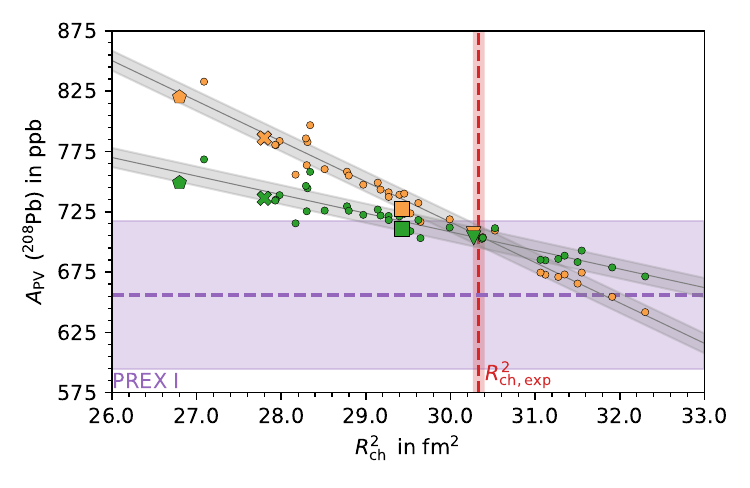}
    \caption{Correlation of \apv{} with \rchsq{} for $Q_\text{weak}$~\cite{Qweak:2021ijt} (upper panel) and PREX I~\cite{Abrahamyan:2012gp} (lower panel), analogous to Fig.~\ref{fig:corr_APV_main}.}
    \label{fig:corr_APV_Qweak}
\end{figure}

We evaluate the expectation values of radius operators~\cite{Heinz:2024juw}
and the nuclear structure functions required to construct the charge and weak densities~\cite{Hoferichter:2020osn}.
We use the Gaussian factorization of the center-of-mass wave function to make our densities translationally invariant~\cite{Hagen:2009pq, Heinz:2024cwg, Miyagi:2025rvx, He:2026zie}.
From this, we obtain the charge radius $\rch$, the point-proton and point-neutron radii $\rprot$ and $\rneut$, the neutron skin \rskin{},
and the full intrinsic charge and weak densities
$\rho_\mathrm{ch}$ and $\rho_\mathrm{w}$ of \elem{Ca}{48} and \elem{Pb}{208} for each Hamiltonian.

Our ensemble of Hamiltonians explores the EFT truncation uncertainty, but our many-body calculations are also approximate.
Our many-body computations are expanded in a truncated basis.
The effect of varying this truncation was explored in Refs.~\cite{Heinz:2024cwg, Miyagi:2025rvx},
and in Fig.~\ref{fig:corr_truncation}
we show the impact on predicted radii of \elem{Pb}{208} when varying our basis truncation from 13 major shells ($e_\mathrm{max}=12$) to 15 major shells ($e_\mathrm{max}=14$).
The radii shift by small amounts along the correlations established from the ensemble of Hamiltonians.
This indicates that basis truncation uncertainties are correlated in the same way as Hamiltonian uncertainties for these observables.
The IMSRG(2) approximation may be systematically improved by going to the next order, the IMSRG(3)~\cite{Heinz:2021xir, Heinz:2024juw}.
Such computations are too expensive for calculations of \elem{Pb}{208}, but were explored for \elem{Ca}{48} in Ref.~\cite{Heinz:2024cwg}.
Again the corrections going beyond IMSRG(2) followed the established correlations very closely.
For this reason, we assume that basis truncation and many-body method truncation uncertainties are negligible compared to the residual EFT uncertainty captured by the correlation band.

\begin{figure}[t!]
    \centering
    \includegraphics[width=0.95\linewidth,trim= 0 0 0 10,clip]{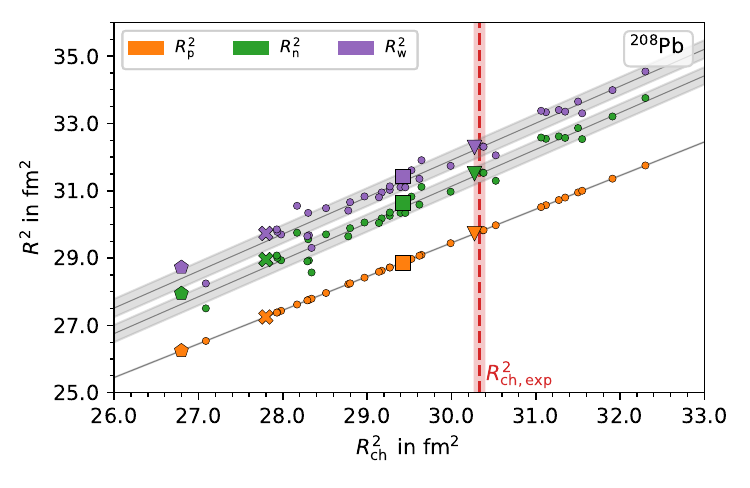} 
    \includegraphics[width=0.95\linewidth,trim= 0 0 0 10,clip]{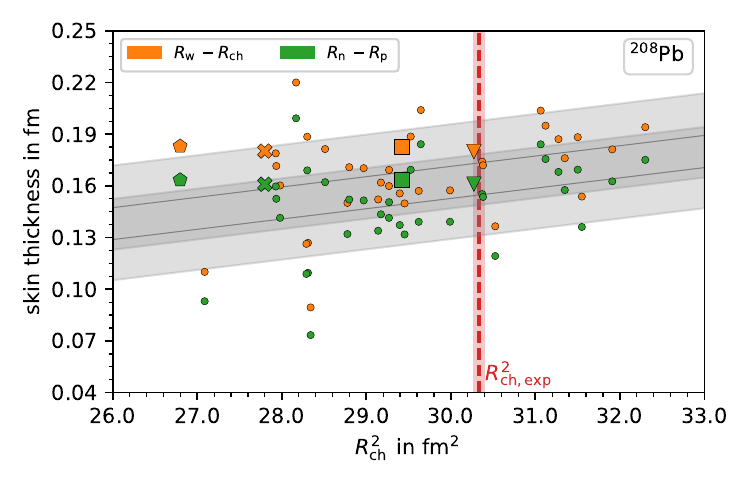}
    \caption{Correlation between radii squared/skin thicknesses (upper/lower panel) and \rchsq{} for \elem{Pb}{208}.}
    \label{fig:corr_rsq_rskin}
\end{figure}

\section{Correlations}
\label{app:further_correlations}

For completeness, we provide explicit values for the correlations given in Fig.~\ref{fig:corr_APV_main}.
The correlations between \apv{} and \rchsq{} are:
\begin{align}
    \mathrm{APV1}:&& \apv(\elem{Ca}{48}) &= -705.2 \, \Delta \rchsq+ 2384 \pm 62\,, \notag\\
    \mathrm{APV2}:&& \apv(\elem{Ca}{48}) &= -176.2 \, \Delta \rchsq + 2387 \pm 82\,, \notag\\
    \mathrm{APV1}:&& \apv(\elem{Pb}{208}) &= -21.52 \, \Delta \rchsq + 570.4 \pm 5.3\,,\notag\\
    \mathrm{APV2}:&& \apv(\elem{Pb}{208}) &= -8.239 \,\Delta \rchsq + 570.1 \pm 5.1\,, 
\end{align}
where $\Delta \rchsq=\rchsq-R^2_\text{ch,exp}$ (in fm$^2$) and \apv{} is given in ppb. 
The correlations between \apv{} and \rskin{} are:
\begin{align}
    \mathrm{APV1}:&& \rskin(\elem{Ca}{48}) &= -27.62 \, \Delta \tilde{A}_\text{PV} + 0.146 \pm 0.016\,,\notag \\
    \mathrm{APV2}:&& \rskin(\elem{Ca}{48}) &= -122.2 \, \Delta \tilde{A}_\text{PV} + 0.120 \pm 0.012\,, \notag\\
    \mathrm{APV1}:&& \rskin(\elem{Pb}{208}) &= -406.2 \, \Delta \tilde{A}_\text{PV} + 0.165 \pm 0.022\,, \notag\\
    \mathrm{APV2}:&& \rskin(\elem{Pb}{208}) &= -1390 \, \Delta \tilde{A}_\text{PV} + 0.187 \pm 0.018\,,
\end{align}
where $\Delta \tilde{A}_\text{PV}= (\apv-A_\text{PV,exp}) \times 10^{-6}$ (with \apv{} in ppb) and \rskin{} is given in fm.

Our main results for the prediction of \apv{} derive from the correlation with \rchsq, as given above and shown in the left panel of Fig.~\ref{fig:corr_APV_main}. Table~\ref{tab:APV_tab} includes the same results for $Q_\text{weak}$~\cite{Qweak:2021ijt} and PREX~I~\cite{Abrahamyan:2012gp}, based on the correlations shown in Fig.~\ref{fig:corr_APV_Qweak}.

Independently of \apv, our ab initio calculations allow us to study the relations between charge, point-proton, point-neutron, and weak radii for the nuclei considered, see also Ref.~\cite{Heinz:2024cwg} for \elem{Ca}{48} and \elem{Al}{27}. The radii squared can then be extracted as the $r^2$-moments of the normalized version of their respective distributions
\begin{align}
    R^2_x &= \frac{4 \pi}{Q_x}\int_0^\infty dr\,r^4 \rho_x(r),  & Q_x & =4 \pi \int_0^\infty dr\, r^2 \rho_x(r),
\end{align}
with $x \in \{p,n,\text{ch},\text{w}\}$ and ${Q_p=Q_\text{ch}=Z}$ and ${Q_n=N}$. Moreover, the point-proton and point-neutron distributions are calculated via Fourier transform of the structure functions of the $M_{L=0}$ multipoles, analogously to the charge and weak distributions. Studying the correlation between the radii over the different input chiral Hamiltonians, we may then infer proton, neutron, and weak charge radii squared by considering their correlation with the charge radius squared. For \elem{Pb}{208}, we obtain the correlations shown in Fig.~\ref{fig:corr_rsq_rskin}, while the corresponding figures for \elem{Ca}{48} and \elem{Al}{27} were already provided in  Ref.~\cite{Heinz:2024cwg}.
The resulting numerical values for \elem{Pb}{208} are summarized in Table~\ref{tab:rsq_rskin}.

\begin{table}[t!]
	\renewcommand{\arraystretch}{1.3} 
	\centering 
    \begin{ruledtabular}
    \begin{tabular}{l r r r}
    & \multicolumn{1}{c}{This work} & \multicolumn{2}{c}{Experiment} \\ \colrule
    \multirow{3}{*}{\rch{}} & \multirow{3}{*}{$5.508(6)$} & $5.4989(7)$ & \cite{DeVries:1987atn} \\
    & & $5.503(2)$ & \cite{DeVries:1987atn,PREX:2021umo} \\
    & & $5.5012(13)$ & \cite{Angeli:2013epw} \\
    \rprot{}  & $5.458(7)(0)$ & \\
    \rneut{}  & $5.612(7)(23)$ & \\
    \rw{}  & $5.681(7)(24)$ & $5.800(75)$ & \cite{Abrahamyan:2012gp,PREX:2021umo} \\
    $\rneut - \rprot$  & $0.155(0)(24)$ & $0.283(71)$ & \cite{Abrahamyan:2012gp,PREX:2021umo} \\
    $\rw - \rch$  & $0.173(0)(24)$ & &\\
	\end{tabular}
    \end{ruledtabular}
    \caption{Radii and skin thickness of \elem{Pb}{208} in fm derived from correlations  with \rchsq{}. The first uncertainty refers to the uncertainty coming from $\rchexp$, the second one to the residual uncertainty in the correlation.} 
	\label{tab:rsq_rskin}
\end{table}

\bibliography{ref}

\end{document}